\documentclass[aps,prl,twocolumn]{revtex4}
\usepackage{color}
\usepackage{epsfig} 

\begin{document}

\title{Coupled  Superconducting Phase and
Ferromagnetic Order Parameter Dynamics }

\author{I. Petkovi\'{c}$^{1}$$^*$, M. Aprili$^{1}$, S.
E. Barnes$^{2,3}$, F. Beuneu$^{4}$, and S. Maekawa$^{5,6}$}

\affiliation{
$^{1}$  Laboratoire de Physique des Solides, Universit\'{e} Paris-Sud,
CNRS, UMR 8502, 91405 Orsay, France. $^{2}$ Theory of Condensed Matter
Group, Cavendish Laboratory,
Madingley Road, Cambridge CB3 0HE, United Kingdom. $^{3}$Physics
Department, University of Miami, Coral Gables, FL 33124, USA.
$^{4}$Laboratoire des Solides Irradi\'{e}s, CNRS, \'{E}cole
Polytechnique, F-91128 Palaiseau, France. $^{5}$Institute for Materials
Research, Tohoku University, Sendai 980-8577, Japan. $^{6}$CREST,
Japan Science and Technology Agency, Sanbancho, Tokyo 102-0075, Japan.}

\altaffiliation {petkovic@lps.u-psud.fr}


\date{\today}

\begin{abstract}
Via a direct coupling  between the magnetic order parameter and
the singlet Josephson supercurrent, we detect  spin-wave resonances, and their dispersion,
 in ferromagnetic Josephson
junctions in which the usual insulating or metallic barrier is
replaced with a weak ferromagnet. The coupling arises
within the Fraunhofer interferential description of the Josephson
effect, because the magnetic layer acts as a time dependent phase
plate. A
spin-wave resonance at a frequency $\omega_s$ implies a dissipation
that is
reflected as a depression in the current-voltage curve of the Josephson
junction when $\hbar \omega_s  = 2eV$.  We have  thereby performed a
resonance experiment on only $10^7$ Ni atoms.
\end{abstract}

\maketitle

The coupled dynamics of the electromagnetic field and a Josephson
junction
has a number of manifestations and is very well understood \cite{Josephson,shapiro,fiske,likharev}.  When the usual insulating or
metallic barrier is
replaced with a weak ferromagnet there is a coupling to another field,
namely the spontaneous magnetisation of the ferromagnet. Spin-waves are
elementary spin excitations which can be viewed as both spatial
and time dependent variations of the magnetisation. In a
ferromagnet the lowest energy excitation, the Ferromagnetic Resonance
(FMR), corresponds to the uniform precession of the magnetisation around
an externally applied magnetic field at the frequency $\omega_s$. This
mode can be resonantly excited by an alternative (ac)
magnetic field that couples directly to the magnetisation, as
described by the  Landau-Lifshitz equations \cite{Kittel}.
 The Josephson phase difference $\phi$ between the  two superconductors  has
its own dynamics. A bias voltage $V_0$ causes $\phi$ to become  time-dependent
so that  $\phi=\phi_0 +
\omega_J t$, where $\omega_J = (2e/\hbar) V_0$ and $\phi_0$
is  arbitrary.   Corresponding to the ac Josephson effect \cite{Josephson}, for  our junctions, to a good approximation, the resulting ac Josephson current density is
$J_s= J_c \sin (\phi_0+\omega_J t)$, where $J_c$ is the critical
current density.

In analogy with the A-phase \cite{3Heref} of $^3$He,  coupled
magnetic  and phase oscillations should exist in ferromagnetic
superconductors with  triplet pairing, but have never been observed.
We show here that a  similar coupling for singlet
superconductors can be realised in a Josephson junction with a
ferromagnetic barrier. The dynamical coupling stems from the spatial
interference of the Aharonov-Bohm phase caused by $\mathbf{M} (t)$,
resulting in the spatial dependence of $\phi(\mathbf{r},t)$. The ac
Josephson current  produces an oscillating magnetic field $\mathbf{H}
(t) $ and when the Josephson frequency matches the spin wave
frequency, $\omega_J\approx \omega_s$, this resonantly excites $
\mathbf{M}(t)$. Due to the nonlinearity of the Josephson effect,
there is a rectification of current across the junction, resulting
in a dip in the average  dc component of $J_s$ at voltage $V_s=(\hbar/
2 e)\,
\omega_s$. The principal result reported here is the observation of
these coupled dynamics.

Magnetised Josephson junctions \cite{Kontos} require weak ferromagnetic
materials and nanosized junction area to keep the overall magnetic
flux in the junction smaller than the flux quantum $\Phi_{0}$.
An electron microscope image of a typical ferromagnetic junction
used in this study is shown in Fig.~1(a), while
Fig.~1(b) is a schematic representation of the different layers. The
superconducting electrodes comprise 50~nm of Nb ($T_c=7.6$~K), while the
barrier is 20~nm of Pd$_{0.9}$Ni$_{0.1}$ ($T_{\rm Curie}=150$~K).
The current-voltage (IV) characteristics are measured using current
bias and are reported, as function of the applied in-plane field, in the
right insert of Fig.~$2$.
The IV characteristics are not hysteretic,
and overall they correspond closely to those expected for a junction
with a conductive barrier \cite{stewart,mccumber}. The junction normal
resistance is $R_n \approx 0.8\, \Omega$, and the Josephson coupling
is $I_c R_n\! \approx\! 5\,\mu$V, as expected for ferromagnetic
junctions of this thickness \cite{Kontos}, yielding the critical
supercurrent of $I_c\!\approx\! 7 \,\mu$A. The hostile nature of even
a weak
ferromagnetic environment for singlet Cooper pairs is illustrated by a
similar junction with 70 nm of nonmagnetic Pd which, despite the
almost four times larger thickness, has a larger critical current $I_c \approx 44 \,\mu$A.

\begin{figure}[h]
\centerline{\hbox{  \epsfig{figure=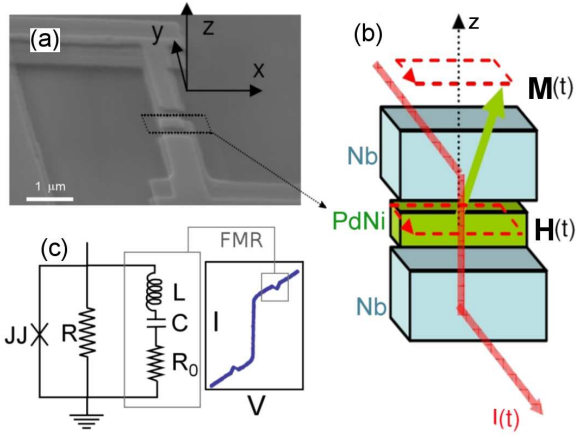,width=80mm}}}
\caption{(color online). (a) SEM photo of the
ferromagnetic Josephson junction used in the experiment. Junction area
is $\rm 500 \,nm
\times 500$~nm. (b) Schematic cross-section of the Josephson junction.
The ac Josephson current $I(t)$ flows through the
junction creating
an rf magnetic field $\mathbf{H}(t)$, causing the magnetisation
precession $\mathbf{M}(t)$, which in turn resonantly couples with the
Josephson phase at frequency $\omega_J$. Layers are respectively from
the bottom: Nb ($50$~nm), $\rm
Pd_{0.9}Ni_{0.1}$ ($20$~nm) and Nb ($50$~nm).  (c) The
equivalent RSJ model and the sketch of the effect of the FMR on the
current-voltage characteristics. The ferromagnetic layer is modeled as
a series LCR$_0$ oscillator, in parallel with the Josephson
junction.  Resistance $R_0$ is proportional to the imaginary part of
the susceptibility $
\chi''$.}
\end{figure}

For a square junction of side $L$, the total supercurrent is
given by  the integral \cite{barone}
\begin{equation}
\label{current}
I_s = J_c \int_{-L/2}^{L/2} dx\int_{-L/2}^{L/2} dy \sin \phi(x,y,t)
\label{une1}
\end{equation}
with
\vspace{-5pt}
\begin{equation}
\label{phase}
\phi(\mathbf{r},t)  = \phi_0 +  \omega_J t  - \frac{2e}{\hbar} \int
\mathbf{A}\cdot
d \mathbf{r},
\label{une2}
\end{equation}
where  the last term  is the
Aharonov--Bohm phase \cite{rowell}, involving the vector potential $
\mathbf{A}$. We use a
gauge where $\mathbf{A} = A(\mathbf{r},t)\, \mathbf{\hat{ z}}$, the
direction $\mathbf{\hat{z}}$
being perpendicular to the junction
surface [see Fig.~1]. Therefore $\phi(\mathbf{r},t) = \phi_0 + kx +
\omega_J t +\phi_m$, where $\phi_m = (4a e/\hbar) A_{mz}$  reflects
time dependent fields and $k = (4ed/\hbar) \mu_0 H + (4ea/\hbar) \mu_0
M_{0y}$.  Here $M_{0y}$ is the $y$ component of the static
magnetisation $
\mathbf{M_0}$, the applied field $\mathbf{H}$ is in the $y$ direction,
$2a$ and
$2d\!=\!2(a+\lambda)$ are the actual and magnetic thickness of barrier
and
$\lambda$ the London penetration depth. Equations (\ref{une1}) and
(\ref{une2}) are
used to describe both the statics and the dynamics of our junctions.

In the absence of a bias $(V_0 = 0)$, we are dealing with static fields, and the Equations
(\ref{une1}) and (\ref{une2}) lead to the Fraunhofer  pattern $I_s =
J_c L\int_{-L/2}^{L/2}dx \sin k x$ \cite{rowell}.
The
magnetisation $
\mathbf{M_0}$ of the barrier has the same effect as inserting a wedge
shaped phase plate in front of the slit, it displaces the diffraction
pattern.
 Experimentally, the diffraction  pattern is shifted to the right for increasing  (positive $M_{0y}$), and
 the left for  decreasing  (negative $M_{0y}$), fields. This illustrates the
linear
nature of  the coupling to $\mathbf{ M}$.  In  Fig.~2,  the dotted curves are a fit using Eqs.~(\ref{une1})
and (\ref{une2}), along with the
magnetisation data measured on a trilayer with the same cross section
as the junction [see the left insert of Fig.~2]. The periodicity and
the asymptotic behaviour of the measured diffraction pattern attest to
the high quality of our junctions. They confirm the close-to-uniform
current
distribution and single-harmonic current-phase relation, while the
reproduction of the shift with the two sweep directions, using
experimental
magnetostatic data, confirms the validity of our description.

\begin{figure}[t]
\centerline{\hbox{  \epsfig{figure=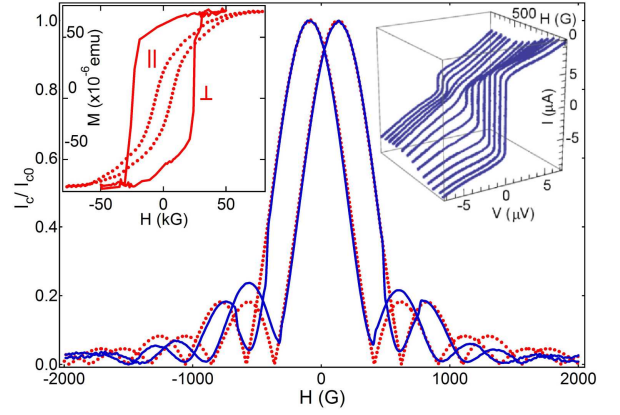,width=80mm}}}
\caption{(color online) Dependence of the Josephson
critical current on the external
magnetic field in plane. The solid curve represents the normalized
experimental data taken at 35~mK and the dotted curve is the Fraunhofer
pattern  expected for our junction parameters including the
magnetisation. Left insert: The magnetic hysteresis taken at 10~K on a
millimetric trilayer with the same cross section as the junction, with
field applied in plane (dotted curve) and perpendicular to the plane
(solid curve). Right insert:
Current-voltage curves for different field values, measured at 35~mK.
}
\end{figure}

The dynamical coupling  reflects a similar
phase contribution $\varphi(\mathbf{r},t)$ due to $\mathbf{ M}
(\mathbf{r},t)$, but which now has both a temporal and
a spatial dependence, the equivalent of a phase plate in the optical
analog with a
similarly dependent refractive index $n(\mathbf{r},t)$. The
dynamics of the magnetisation (timescale of $1$ ns) is
much slower
than the diffusion time through the ferromagnetic layer ($0.5$ ps).
The Josephson coupling is thus adiabatic with respect to the
magnetisation dynamics. This assumption is implicit in
Eqs.~(\ref{une1}) and (\ref{une2}).  The signal is seen for  $V>I_c R
$, implying Eq.~({\ref{une1}) can be linearized. The dc magnetic
signal then corresponds to \cite{barone}
\begin{equation}
\label{trois1}
I_m =  \frac{4ae}{\hbar} \int_{-L/2}^{L/2}dx  \int_{-L/2}^{L/2} dy J_c
\overline{ \cos ( kx + \omega_J t) \phi_m}\, ,
\end{equation}
where the bar denotes a time average. Substituting for  $
\phi_m = (4a e/\hbar) A_{mz}$  and using $\mathbf{J} = \mathbf{\nabla}
\times \mathbf{H}$, following both a time and space integration by
parts, the dc signal reflecting the magnetic resonance is
\begin{equation}
\label{trois2}
I_m =
\frac{1}{ V_0} \int d\mathbf{ r}\, \overline{\mathbf{ H}\cdot \frac{d
\mathbf{ M}}{dt}},
\end{equation}
with $M_i (\mathbf{ r},t) =
\int dt^\prime \chi_i(t-t^\prime) H_i(\mathbf{ r},t^\prime)$, $i = x,y,z
$, where $\chi_i(t) $ is the dynamic susceptibility. This has an
appealing interpretation in terms of magnetic losses. Here, as
illustrated by Fig.~1(b),  $\mathbf{ H}(\mathbf{ r},t)$ is the magnetic
field
which circulates  inside the junction by virtue of the ac Josephson
current. The junction lateral size $L$ is smaller than both $\lambda$
and the skin depth for the frequencies involved. The displacement
current is therefore negligible and all that is needed is to integrate
Amp\`ere's law in order  to determine  $\mathbf{ H}(\mathbf{ r},t)$.
More details of these calculations are given elsewhere \cite{barnes}.
The  current due to $\mathbf{ M} (\mathbf{r},t)$ is
\begin{equation}
\label{trois4}
I_m =2 \pi I_c (0) \frac{ \Phi_{\rm rf}}{\Phi_0}\left[F_x\chi^{\prime
\prime}_x
(\omega_J)
+ F_y\chi^{\prime\prime}_y (\omega_J)\right],
\end{equation}
where $\Phi_{\rm rf} = (2aL) B_{\rm rf} = (2aL)  \mu_0 I_c(0)/
L$ is the flux due to the radio frequency field and $F_x = (1/12)
(I_c(B_0)/I_c)^2$ and $F_y = (2/\theta_L^2)[1+\sin (\theta_L/2) \cos (\theta_L/2) +
((13/12) - (4/\theta_L^2) )\sin^2 (\theta_L/2)]$;
$\theta_L=kL$,  reflect the geometrical structure of the coupling. As the
equilibrium magnetisation is along the $z$ axis,
the magnetic resonance
signal is contained in $\chi^{\prime\prime}_x
(\omega_J)$ and $\chi^{\prime\prime}_y (\omega_J)$,  the
Fourier transforms of the imaginary part of the susceptibility.
Therefore, the  total dc current within the Resistively Shunted
Josephson junction (RSJ) model \cite{stewart,mccumber} is
\begin{equation}
\label{30}
I=  \frac{V_0}{R(0)} + \frac{{I_c}^2(B_0)}{2V_0} R(\omega_J)-I_m,
\label{rsj}
\end{equation}

 \noindent where $R(0)$ and $R(\omega_J) \approx R$ are the junction resistances for dc and frequency $\omega_J$. A simple physical
argument can account for the three terms in
Eq.~(\ref{rsj}). The average power dissipated in the junction is
$IV_0$ and so the first term, ${V_0}^2/R$, corresponds to  the Ohmic
loss at dc, while  $\frac{1}{2}{I_c}^2R(\omega_J) $  is the similar
loss at $\omega_J$. The
key third term represents a self-inductance $L(\mathbf{ M})$, stemming
from the ferromagnet, in parallel  with the junction
and modeled as an
LCR$_0$ oscillator [see Fig~1(c)], where $R_0$ reflects the magnetic damping. At  the  magnetic resonance frequency,
energy is absorbed by the ferromagnet, causing the oscillator to be
lossy. This actually reduces the
effective junction resistance, leading to a dip in $I(V)$. In this
manner, the
Josephson junction rectifies the self-induced magnetic resonance.

\begin{figure}[h]
\vspace{2mm}
\centerline{\hbox{  \epsfig{figure=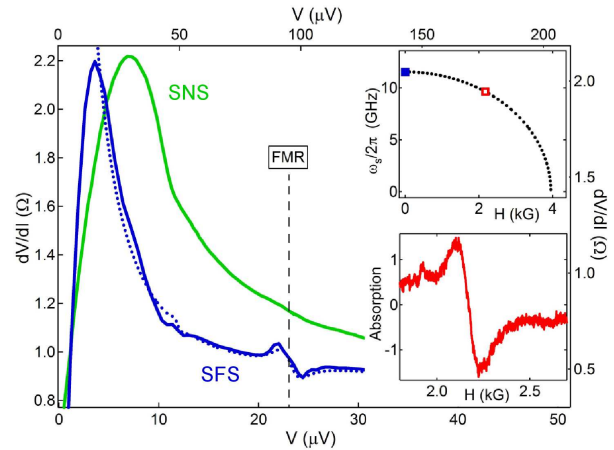,width=80mm}}}
\caption{(color online). Dynamical resistance of the ferromagnetic
Josephson junction (solid curve, SFS, bottom and left
axis) shows resonances compared to a similar non-ferromagnetic
junction (solid curve, SNS, top and right
axis). Dotted curve is a fit to theory [Eq. (\ref{rsj})]. The mode
at 23~$\mu V$ is the ferromagnetic resonance (FMR).
Bottom insert: Conventional cavity ferromagnetic resonance on a
macroscopic trilayer. Top insert: Comparison between the field
dependence of the FMR in the Josephson junction (solid square) with the
cavity measurement on a macroscopic trilayer (open square). Dotted curve
is a parameter free fit of the FMR using the Kittel formula [Eq.
(\ref{150})].}
\end{figure}

This coupling to the magnetic system is evident in the measured
dynamical
resistance $dV/dI$ curves reported in Fig.~3. We measure the dynamical
resistance rather than the IV characteristics to improve amplitude
resolution. The mode labeled FMR (Ferromagnetic Resonance) is seen
only for ferromagnetic junctions.
There is good agreement between the experiment, solid
curves, and theory, dotted curves.
The magnetic resonance mode observed in our experiments reflects the
properties of a thin film of the ferromagnet Pd$_{0.9}$Ni$_{0.1}$.
Magnetisation curves $M(H)$, measured directly for a large area $\rm
Nb/PdNi/Nb$ trilayer with the same cross section as the
junction, are shown in the insert of Fig.~2. They
indicate that $\mathbf{M}$  is perpendicular to the junction plane, a
conclusion
reinforced by
earlier anomalous Hall effect measurements on similar thin
films \cite{kontos2001}. The FMR mode, shown in  Fig.~3,  occurs
at $V_0 =
23\,\mu$V. This is unambiguously identified as such, since the
frequency
$\omega_s=2eV_0/\hbar$
agrees, without fitting parameters, with  the
Kittel formula \cite{Kittel1}

\vspace{-5pt}
\begin{equation}
\label{150}
\omega_s = \gamma_e \sqrt{(H_K - 4 \pi M_S)^2 - H^2}
\end{equation}
for the in-plane magnetic field dependence of the uniform FMR mode
when
the anisotropy field is perpendicular to the plane.  The anisotropy
field
$H_K= 4900$ G and the magnetisation at saturation $M_{S}=930$ G are
both
determined directly from the  static magnetisation data, and
$\gamma_e=\mu_B/\hbar$, where $\mu_B$ is the Bohr magneton. For
comparison,  the ferromagnetic resonance of a
macroscopic Nb/PdNi/Nb trilayer has been measured in a conventional
$9.5$ GHz cavity
spectrometer at
$10$ K with field applied parallel to the substrate. The cavity FMR,
shown in the bottom
insert of Fig.~3,
occurs at $2160$~G,  again exactly as predicted by Eq.~(\ref{150}).
Displayed in the top insert of Fig.~3 is the comparison of the
resonant
mode in the Josephson junction (solid square) and in the macroscopic
trilayer (open square). The dotted curve shows  the frequency
of the FMR mode
calculated  from the
Kittel formula [Eq. (\ref{150})]. The spectra presented in the main
part of Fig.~3 contain an extrinsic
broadening caused by a lock-in modulation voltage of $\sim 1\,\mu$V.
For
the ac modulation voltage of 0.5 $\mu$V, the junction resonance width
saturates at 0.5 $\mu$V, which corresponds to the conventional
resonance
width (150~G). The signal amplitude corresponds to a
resonant susceptibility of approximately 10, consistent with the FMR
mode measured in a microwave cavity and reported in the bottom insert
of Fig.~3.

\begin{figure}[h]
\vspace{2mm}
\centerline{\hbox{  \epsfig{figure=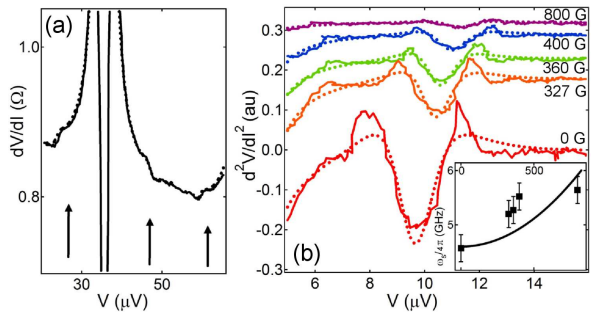,width=80mm}}}
\caption{(color online). (a) The
dynamical resistance of the ferromagnetic Josephson junction with
applied external microwaves. Pronounced dip is the Shapiro
resonance, and arrows indicate sideband resonances at the same
frequency as the  $n\!=\!2$ mode from Fig.~3. The external
microwave frequency is 17.35~GHz  and the temperature 35~mK.  (The solid curves
are
the measured derivative of the dynamical resistance from Fig.~3, and
dotted curves are a fit [Eq. (\ref{rsj})]. The spin wave
frequency increases with field. Insert: The   field dispersion
relation of the
modes. The solid curve is Eq.~(7) when the spatial
dependence of the FMR modes is taken into account.}
\end{figure}

In order to demonstrate that the magnetic system is coupled to the
super-
but not to
the normal current, we have performed  Shapiro step \cite{shapiro}
measurements, reported as the dynamical resistance $dV/dI$ in Fig.~4(a).
The
junction is irradiated with microwaves of frequency  $\nu\!=\!
17.35$~GHz at 35~mK. The Shapiro steps arise from the mixing of the
microwave signal with the ac Josephson effect and are smaller replicas
of  the
zero-voltage current step displaced from zero voltage by $V_{n}
(\hbar/2e)\, \omega$, where $\omega\! = \! 2\pi \nu$, and $n$ is an
integer. We do not observe half integer Shapiro steps, indicating
negligible higher harmonics in the current-phase relation. However, as
expected within the RSJ model, the ferromagnetic resonance can be
exited at voltage $V_{ns}\!=\!(\hbar/2e)\,\omega_s/n$ \cite{likharev}.
The sub-harmonic for $n\!=\!2$ is visible in the spectrum in Fig.~3.
As shown in Fig.~4(a), it is reproduced as a side-band to each regular
step when $V \!=\! (\hbar/2e)(n
\omega \pm \omega_{\rm s}/2)$. Experimentally, we do not have available
a high enough
frequency to separate similar side-bands for the main FMR mode at $
\omega_s$.

Finally, the field dependence of the resonance at $\omega_s/2$
has been studied in more detail in the  second derivative, $d^2V/dI^2$
[Fig.~4(b)], where the minima correspond to $V_{2s}=(2e/\hbar)\,\omega_s/
2$. Measurements were limited in
field due to  the rapid decrease of the critical current above 800~G.

In the insert of Fig.~4(b), we show  $(1/2\pi)\,\omega_s/2$ as a function
of the applied magnetic field. The error bars are due to the drift of
the amplifier. }The solid curve is Eq.~(\ref{150}), without fitting
parameters, with the spatial dependence of FMR  taken into
account. The spatial dependence of the spin-waves
leads to an additional term  to
Eq.~(\ref{150}) given by $ a k^2$, where  $  k \!=\! (\pi d/\Phi_0) H$
is
the spin wave momentum and $ a\!=\!E_{ex}b^2$, where $E_{ex}\!=\!50$~meV
is the PdNi exchange energy and $b\!=\!0.1$~nm the lattice constant.
Since the width of the junction is only about 500~nm, this leads to a
small but finite correction to the uniform FMR energy  which is larger than the direct effect of the applied dc field.  Illustrated in this manner is  the direct determination of spin-wave dispersion using the present technique.

In conclusion, we have demonstrated the dynamical
coupling of the superconducting phase with the spin waves in a
ferromagnet  and measured their dispersion. We have performed a photon free FMR experiment on about
10$^7$ Ni atoms, which would be  infeasible with standard FMR techniques,
and have illustrated a new methodology for the  study of spin
dynamics. There are direct and
implied applications to spintronics and nanomagnetism \cite{spinbook}.

We thank J. Gabelli, B. Reulet, D.
Feinberg, R. Melin, Z. Radovi\'{c}, I. Martin and M. Houzet for
stimulating discussions. M.A. is indebted to H. Bouchiat for an
illuminating conversation
and to A. Thiaville and H. Hurdequint for many tutorials about spin
dynamics.
This work was in part supported by CREST of JST, and  EPSRC(UK).

\end{document}